\pdfoutput=1
\documentclass[11pt]{article}

\usepackage[preprint]{coling}

\usepackage{times}
\usepackage{latexsym}

\usepackage[T1]{fontenc}
\usepackage[utf8]{inputenc}

\usepackage{microtype}

\usepackage{inconsolata}

\usepackage{graphicx}
\usepackage{booktabs}

\title{ES-KT-24: A Multimodal Knowledge Tracing Benchmark Dataset with Educational Game Playing Video and Synthetic Text Generation}

\author{Dohee Kim${^1}{^*}$, Unggi Lee${^1}$${^,}$${^2}{^*}{^\dagger}$, Sookbun Lee${^1}{^*}$, Jiyeong Bae${^1}$, Taekyung Ahn${^1}$\\ 
{\bf Jaekwon Park${^1}$, Gunho Lee${^1}$, Hyeoncheol Kim${^2}{^\dagger}$} \\
Enuma, Inc.${^1}$, Korea University${^2}$ \\
\texttt{\{dohee, unggi, blackdew, jiyoung, taekyung, jaekwon, gunho\}@enuma.com}, \\
\texttt{harrykim@korea.ac.kr} \\
${^*}$ is first authors and $\dagger$ is corresponding authors. \\
}

\begin{document}
\maketitle

\begin{abstract}
This paper introduces ES-KT-24, a novel multimodal Knowledge Tracing (KT) dataset for intelligent tutoring systems in educational game contexts. Although KT is crucial in adaptive learning, existing datasets often lack game-based and multimodal elements. ES-KT-24 addresses these limitations by incorporating educational game-playing videos, synthetically generated question text, and detailed game logs. The dataset covers Mathematics, English, Indonesian, and Malaysian subjects, emphasizing diversity and including non-English content.

The synthetic text component, generated using a large language model, encompasses 28 distinct knowledge concepts and 182 questions, featuring 15,032 users and 7,782,928 interactions. Our benchmark experiments demonstrate the dataset's utility for KT research by comparing Deep learning-based KT models with Language Model-based Knowledge Tracing (LKT) approaches. Notably, LKT models showed slightly higher performance than traditional DKT models, highlighting the potential of language model-based approaches in this field.

Furthermore, ES-KT-24 has the potential to significantly advance research in multimodal KT models and learning analytics. By integrating game-playing videos and detailed game logs, this dataset offers a unique approach to dissecting student learning patterns through advanced data analysis and machine-learning techniques. It has the potential to unearth new insights into the learning process and inspire further exploration in the field.
\end{abstract}

\section{Introduction}

\begin{figure}[hbt]
\centering
\includegraphics[width=1\linewidth]{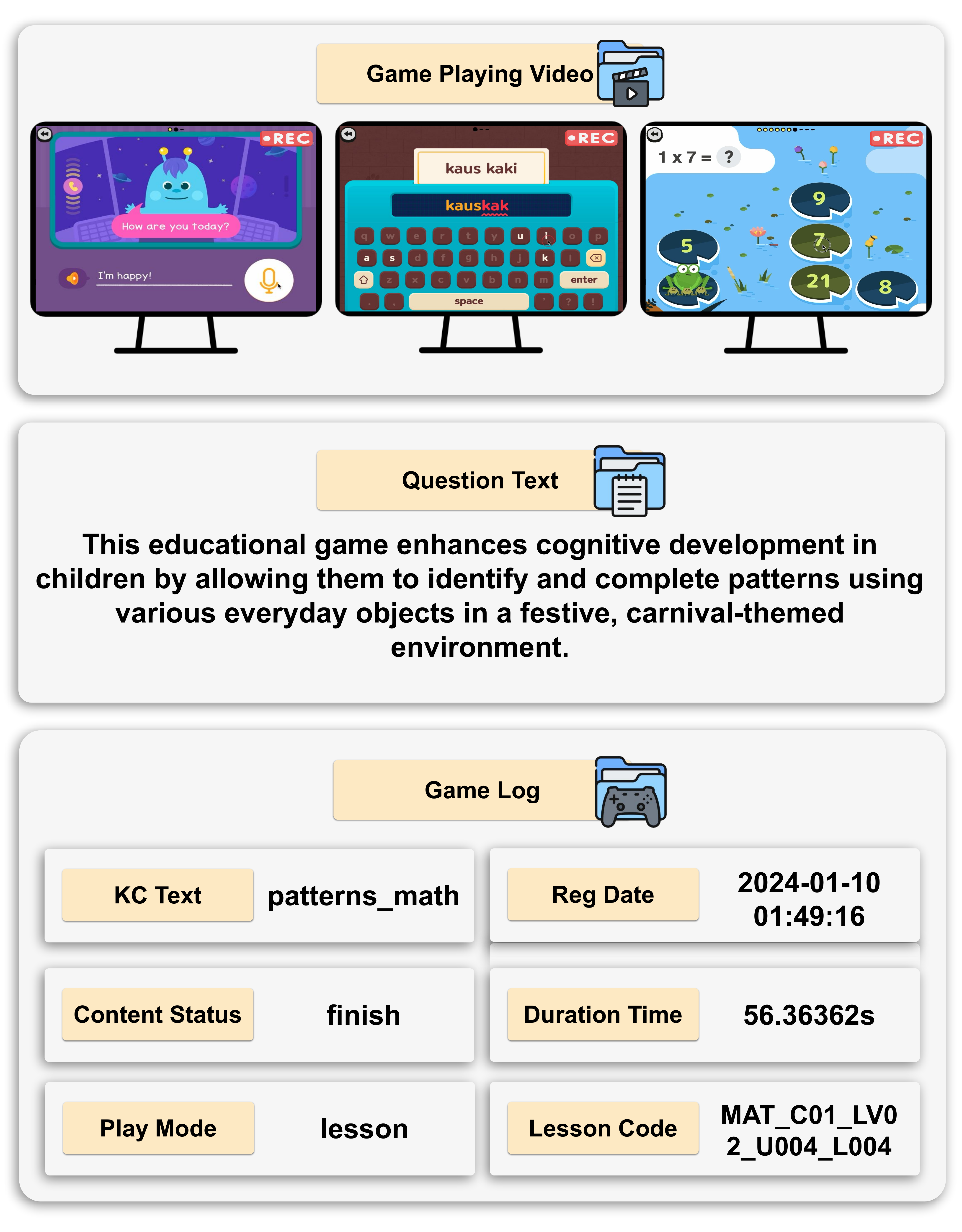}
\caption{
An example of ES-KT-24. ES-KT-24 consists of a multimodal dataset, including game-playing video, synthetic question text, knowledge concept (KC) text, and game logs collected from educational game contexts. 
}
\label{fig:data_info}
\end{figure}

Knowledge Tracing (KT) is a fundamental task that aims to model students' knowledge states over time based on their interactions with learning materials \cite{piech2015deep}. These interactions typically include viewing problems, attempting solutions, and selecting answers in online learning systems. The goal of the KT model is to use these sequences to predict students' future performance on unseen items \cite{lee2024language}. Over the years, various KT models have been developed, ranging from traditional approaches like Bayesian Knowledge Tracing (BKT) \cite{corbett1994knowledge} to more recent deep learning-based methods such as Deep Knowledge Tracing (DKT) \cite{piech2015deep}. These models have shown promising results in predicting student performance and understanding learning patterns.

\begin{figure*}[hbt!]
    \centering
    \includegraphics[width=1\linewidth]{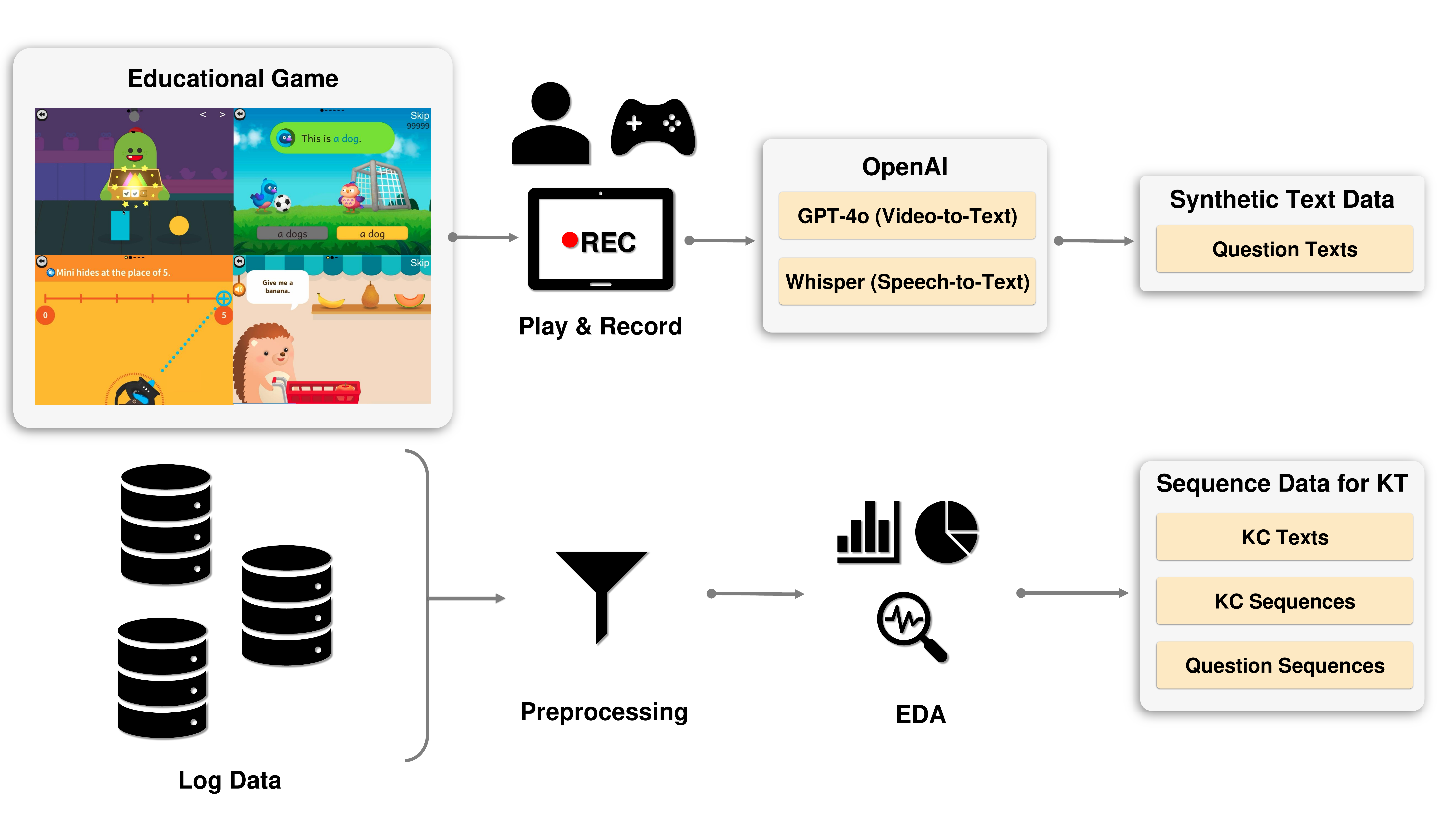}
    \caption{
    The process began with the manual game-play of educational games, which were screen-recorded. These recordings were then converted to text using OpenAI GPT-4o \cite{openai2024gpt4o} for visual content and Whisper \cite{pmlr-v202-radford23a} for audio transcription. The resulting text was utilized to create Questions corresponding to each game. Student problem-solving histories and game logs were preprocessed and explored through Exploratory Data Analysis (EDA), then transformed into sequence data suitable for KT tasks. Finally, this text and sequence data were released as a paired dataset.
    }
    \label{fig:data_gen_flow}
\end{figure*}

However, most existing KT datasets primarily consist of numerical sequences, with only a few recent datasets incorporating textual information. These datasets typically focus on text-based interactions where students submit answers (correct or incorrect) to given problems \cite{lee2024language, lee2024prediction, liu2019ekt}. This approach, while applicable, is limited in its ability to capture the complex interactions and outcomes possible in modern digital educational materials, particularly in game-based learning environments.

Game-based content offers a richer set of interactions and outcomes beyond simple correct/incorrect answers, including metrics such as time spent on tasks or the number of attempts made \cite{hooshyar2022gamedkt}. Game-based learning often involves multiple modalities, including text, images, animations, and audio. To advance research in this area and develop more comprehensive KT models, there is a need for datasets that capture these multimodal aspects of game-based learning.

However, there are still significant limitations in the publicly available datasets \cite{liu2024xes3g5m}. One notable gap is the need for more game-based learning datasets. While educational games have become increasingly popular as learning tools \cite{noemi2014educational, moreno2008educational, petri2024meega+}, KT datasets from gaming environments, mainly publicly available, remain scarce. This scarcity is even more pronounced in game-based learning. To the best of our knowledge, there are currently no publicly accessible KT datasets derived from educational gaming environments.

Another critical limitation is the absence of multimodal data in most KT datasets. There has been a growing trend in research towards multimodal models that combine language models with other modalities such as vision and audio \cite{liu2024visual, deshmukh2023pengi, borsos2023audiolm, driess2023palm}. These multimodal approaches have shown remarkable capabilities in understanding and generating content across different modalities. 

\begin{table*}[hbt!]
\centering
\scriptsize
\setlength{\tabcolsep}{5pt}
\begin{tabular*}{\textwidth}{lrrrrcccc}
\toprule
Dataset & \multicolumn{1}{c}{\# Students} & \multicolumn{1}{c}{\# Questions}  & \multicolumn{1}{c}{\# KCs}  & \multicolumn{1}{c}{\# Interactions}  & \multicolumn{1}{c}{Subjects}  & \multicolumn{1}{c}{Question Texts} & \multicolumn{1}{c}{Videos} &  \multicolumn{1}{c}{Game Logs} \\
% & \multicolumn{1}{c}{Timestamp Avail.}  & \multicolumn{1}{c}{Content Avail.}  & \multicolumn{1}{c}{KC Relation}  & \multicolumn{1}{c}{Ques. Type}  & \multicolumn{1}{c}{Ques. Analysis} \\
\toprule
ASSISTments2009 & 4,217 & 26,688 & 123 & 346,860 & Math & No & No & No \\ %& No & No & No & No & No \\
ASSISTments2012 & 46,674 & 179,999 & 265 & 6,123,270 & Math & No & No & No \\ %& Yes & No & No & No & No \\
ASSISTments2015 & 19,917 & 100 & - & 708,631 & Math & No & No & No \\ %& No & No & No & No & No \\
ASSISTments2017 & 1,709 & 3,162 & 102 & 942,816 & Math & No & No & No \\ %& Yes & No & No & No & No \\
Statistics2011 & 333 & 1,224 & - & 194,947 & Math & No & No & No \\ %& Yes & No & No & No & No \\
Junyi2015 & 247,606 & 722 & 41 & 25,925,922 & Math & No & No & No \\ %& Yes & Yes & No & No & No \\
KDD2005 & 574 & 210,710 & 112 & 809,694 & Math & No & No & No \\ %& Yes & No & No & No & No \\
KDD2006 & 1,146 & 207,856 & 493 & 3,679,199 & Math & No & No & No \\ %& Yes & No & No & No & No \\
NeurIPS2020 & 4,918 & 948 & 57 & 1,382,727 & Math & No & No & No \\ %& Yes & No & No & No & No \\
POJ & 22,916 & 2,750 & - & 996,240 & PL & No & No & No \\ %& Yes & No & No & No & No \\
EdNet & 1,677,583 & 52,676 & 962 & 372,366,720 & Linguistics & No & No & No \\ %& Yes & No & No & No & No \\
DBE-KT22 & 1,361 & 212 & 98 & 167,222 & Computer and Information Science & \textbf{Yes} & No & No \\
XES3G5M & 18,066 & 7,652 & 865 & 5,549,635 & Math & \textbf{Yes} & No & No \\ %& Yes & Yes & Yes & Yes & Yes \\
\textbf{ES-KT-24 (Ours)} & 15,032 & 182 \textbf{(game)} & 28 & 7,783,466 & English, Math, \textbf{Indonesian}, \textbf{Malay}  & \textbf{Yes} & \textbf{Yes} & \textbf{Yes} \\ %& Yes & Yes & Yes & Yes & Yes \\
\toprule
\end{tabular*}
\caption{Comparison of KT benchmark dataset, referencing the XES3G5M research \cite{liu2024xes3g5m}. In several key aspects, existing KT datasets differ significantly from ES-KT-24. ES-KT-24 stands out for its multimodal approach, incorporating text, video, and game log data with educational game context. This dataset also broadens the scope of subjects beyond the typical Math and English, including Indonesian and Malay languages, to promote equity in educational research. Note that the number of questions corresponds directly to the number of games.}
\label{tb:comparison}
\end{table*}

It is becoming increasingly apparent in education that leveraging multiple modalities is crucial for developing comprehensive models that can fully capture the complexity of learning environments \cite{lee2024llava, lee2024see}. Educational contexts often involve rich, multimodal interactions, including visual aids, audio explanations, and hands-on activities, which a sequence of the number cannot adequately represent \cite{blikstein2013multimodal, ochoa2017multimodal, blikstein2016multimodal}.

This paper introduces ES-KT-24 \footnote{The dataset will be made publicly available following final review.}, a novel KT dataset based on educational game data. This dataset is uniquely rich in multimodal features, capturing various aspects of game-based learning environments. As shown in Figure \ref{fig:data_info}, we provide \textit{Game Playing Videos}, where one video was recorded for each available game by researchers, along with the processed \textit{Question Texts} derived from these videos. Additionally, ES-KT-24 also contains \textit{Game Log} such as the game category (\textit{KC Text}), log registration time (\textit{Reg Date}), gameplay status (\textit{Content Status}), gameplay duration (\textit{Duration Time}), gameplay mode (\textit{Play Mode}), and the curriculum code (\textit{Lesson Code}). By providing this resource, we aim to support KT research and broader studies in Learning Analytics (LA) \cite{lang2022handbook}. Researchers can utilize this dataset to investigate game-related learning phenomena, analyze multimodal learning processes, and develop more sophisticated models that account for the diverse modalities present in educational settings. The ES-KT-24 dataset thus represents a significant step towards more comprehensive and realistic modeling of student learning in digital environments. 

Our research contributions are below:

\begin{itemize}
    \item We provide a benchmark for KT using game data, offering a new standard for evaluating KT models in game-based learning environments.
    \item By providing gameplay videos and learning data, we enable researchers in learning analytics to explore and analyze multimodal educational interactions.
\end{itemize}

\section{Related Work}

\subsection{Knowledge Tracing Datasets}

\begin{figure*}[hbt!]
\centering
\includegraphics[width=1\linewidth]{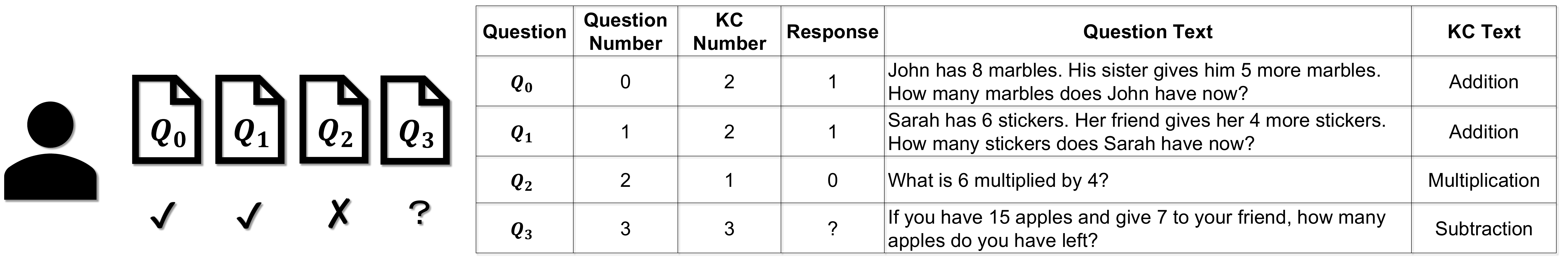}
\caption{
KT dataset format. \textit{Left} shows the exercising process of a student, where the student has already done three questions and will answer question number 4. \textit{Right} shows the corresponding materials of questions that contain their contents and KCs. Note that question and KC texts are used for LKT.
}
\label{fig:kc_def_2}
\end{figure*}

KT primarily involves utilizing datasets that record students' interactions over time \cite{piech2015deep}. These datasets generally contain sequential information about student responses to various items (e.g., KCs or questions). Each interaction in the dataset is typically associated with critical attributes such as the student ID, the item ID, the time the interaction occurred, and whether the student's response was correct \cite{lee2024language} (See Figure \ref{fig:kc_def_2}). By analyzing these sequences, KT models aim to infer a student's underlying knowledge state and predict their future performance on new, unseen items based on patterns in their past interactions \cite{pandey2020rkt, ghosh2020context}.

Several educational datasets have been widely used for KT tasks. These include the ASSISTments datasets \cite{heffernan2014assistments}, which provide instructional guidance by assessing students' knowledge states; the Junyi Academy dataset \cite{JunyiAcademyDataset}, containing mathematics question-solving activities; the Peking Online Judge (POJ) dataset \cite{pandey2020rkt}, offering coding practice data; EdNet \cite{choi2020ednet}, comprising TOEIC test preparation activities; and the Statics2011 \cite{OLI_Engineering_Statics_2011} dataset from an online engineering course. Additionally, data mining competitions have contributed high-quality datasets, such as the KDD Cup 2010 EDM Challenge datasets \cite{KDD_Cup_2010} and the NeurIPS2020 Education Challenge dataset \cite{NeurIPS_2020_Education_Challenge} from Eedi. The KDD Cup 2010 datasets consist of Algebra2005 and 2006 datasets. We refer to Algebra2005 as KDD2005 and Algebra2006 as KDD2006.

While most of these datasets shown in Table \ref{tb:comparison} primarily contain ID features of questions and knowledge components along with timestamps, recent datasets like DBE-KT22 \cite{abdelrahman2022dbe} and XES3G5M \cite{liu2024xes3g5m} stand out by including the actual text content of the questions. XES3G5M, in particular, provides rich auxiliary information, including textual content of questions, knowledge component relationships, question types, and answer analyses that may enhance the modeling process of students' learning outcomes.

However, despite the variety of datasets available, there are two significant gaps in the field of KT datasets. First, KT datasets related to educational games are scarce. While game-based learning environments have become increasingly popular, publicly available KT datasets from these contexts are rare. The GameDKT study \cite{hooshyar2022gamedkt} highlights this gap, presenting one of the few approaches for KT in game-based learning environments, but even this dataset is not publicly available for research. Second, there is a significant lack of KT datasets incorporating multimodal data, including text, images, videos, and audio. The scarcity of multimodal data limits researchers' ability to develop KT models that can leverage diverse types of information to understand and predict student learning processes more accurately.

\subsection{Knowledge Tracing Models}

KT models have evolved significantly, aiming to accurately predict students' knowledge states and future performance. DKT \cite{piech2015deep} significantly advanced the field by leveraging recurrent neural networks to model students' knowledge acquisition over time. Following DKT, various models were developed to enhance performance and address specific challenges. For instance, DKVMN \cite{zhang2017dynamic} utilized a key-value memory network to model the relationships between exercises and knowledge concepts. SAKT \cite{pandey2019self} employed self-attention mechanisms to capture complex dependencies in student interaction sequences. AKT \cite{ghosh2020context} further improved upon this by incorporating a monotonic attention mechanism and context-aware representations.

Recently, a novel approach called LKT has emerged, leveraging the power of Pre-trained Language Models (PLMs) to enhance KT performance. Lee et al. \cite{lee2024language} introduced LKT, which directly utilizes PLMs to process learning data in a textual format. This method has demonstrated superior performance to traditional KT models by effectively capturing semantic information from questions and KCs.

Building upon the success of LKT, efforts are being made to apply it to specific domains. For instance, CodeLKT \cite{lee2024prediction} adapts the LKT framework to programming education, demonstrating significant improvements in predicting student performance on coding tasks.

However, research on KT for game data and multimodal KT has been limited due to a lack of datasets in these areas. In response, this study introduces ES-KT-24, a multimodal dataset specifically designed for KT in game contexts.

\section{Data Description}

\begin{figure*}[hbt!]
\centering
\includegraphics[width=1.1\linewidth]{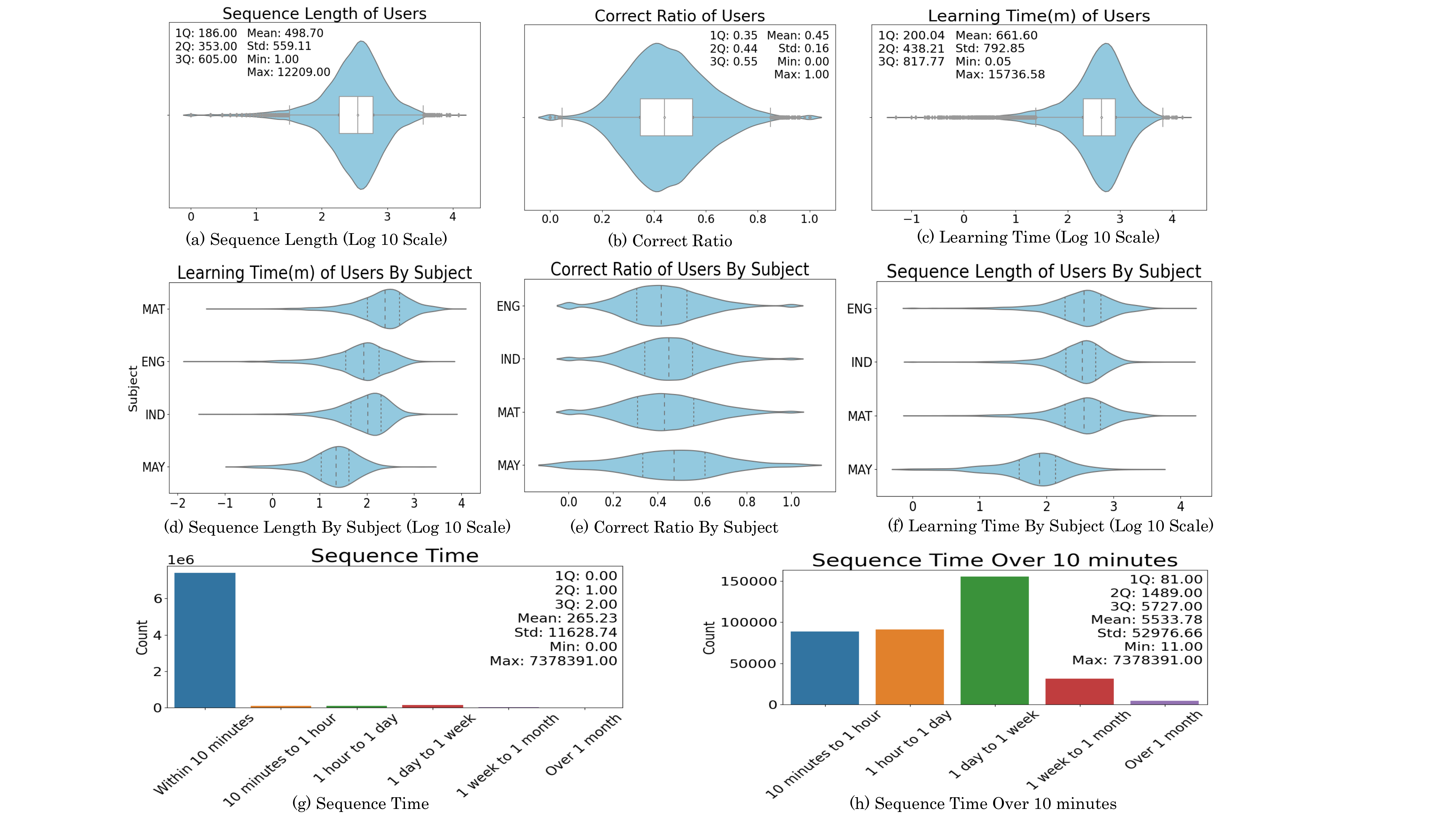}
\caption{
Data analysis results. \textit{Upper} figure (a) to (f) are Violin Plots of Sequence Length, Correct Answer Ratio, and Learning Time. \textit{Lower} figure (g) and (h) are Sequence Time of Users with Subject Comparisons.
}
\label{fig:plot}
\end{figure*}

\subsection{License}

ES-KT-24 is released under the CC BY-NC 4.0 (Attribution-NonCommercial 4.0 International) license. This license restricts the dataset to non-commercial use while allowing modifications and sharing under similar terms. Under this license, others can non-commercially remix, adapt, and build upon the work if they credit the source and indicate if changes were made.

\subsection{Data Collection}

Our data collection process (Figure \ref{fig:data_gen_flow} \textit{left}) involved actual gameplay sessions of educational software comprising game-based content developed by an Edtech company. In this context, a session represents a single playthrough of the game. The dataset consists of logs collected between January 2024 and April 2024. These sessions were screen-recorded to capture the visual aspects of the gameplay. Alongside the video recordings, we collected game information, including player actions, in-game events, correct/incorrect responses, and time duration data. This multimodal approach allowed us to capture both the interactive elements of the game and their corresponding learning outcomes.

The dataset spans four main subjects: Indonesian (IND), Malaysian (MAY), Mathematics (MAT), and English (ENG). Each subject is further divided into specific categories reflecting different educational focuses. These specific categories were used as Knowledge Concepts (KC):

\begin{itemize}
     \item \textbf{Indonesian (IND)}: practice, alphabet, phonics, vocabulary, listening, speaking, writing, test
     \item \textbf{English (ENG)}: alphabet, phonics, vocabulary, listening, reading, speaking, writing
     \item \textbf{Mathematics (MAT)}: numbers, operations, shapes, measurement, data, reasoning
     \item \textbf{Malaysian (MAY)}: practice, alphabet, phonics, vocabulary, speaking, writing
\end{itemize}

Note that the presence of 'practice' and 'test' categories varies across subjects, reflecting differences in educational approaches and assessment methods specific to each subject's curriculum structure and learning objectives.

The dataset comprises 28 distinct KCs and 182 unique content questions, allowing for rich analysis across different learning dimensions. These content questions are textual representations of the educational games created through the process described in Section 3.3, in which each game's content was transformed into a problem or task format.

\subsection{Synthethic Data Generation}

We utilized GPT-4o \cite{openai2024gpt4o}, an advanced language model, to enrich our dataset with textual information. The recorded game-play videos and collected game information were input into GPT-4o to generate synthetic concept texts and problem descriptions. This process allowed us to create a textual representation of the game's educational content, translating visual and interactive elements into descriptive text.

Following the initial generation, we underwent a meticulous review and editing process. This step, which involved manual curation, was pivotal in ensuring the accuracy and relevance of the generated text. We meticulously corrected typographical errors, refined the language for clarity, and adjusted content that did not accurately reflect the game's educational objectives or mechanics. This human-in-the-loop approach was instrumental in maintaining the dataset's quality and educational value while leveraging the capabilities of advanced language models.

\subsection{Data Cleaning and Processing}

Given our educational game, which primarily targets children aged 4 to 6, specific processing considerations were required for the dataset. In this game, users interact through hands-on activities or voice recognition, guided by visual aids and audio explanations. A vital characteristic of the game is that there are no explicit correct or incorrect answers. In other words, to proceed to the next stage of a game, players must input correct answers. This gameplay design limits the data on incorrect attempts, as only correct responses are logged. Consequently, the game logs do not capture correct or incorrect answer data, necessitating a new approach to defining correctness for the KT model. We implemented a rule-based system to infer correctness based on \textit{Content Status} and \textit{Duration Time} in Figure \ref{fig:data_info} to address this.

\begin{figure}[hbt]
\centering
\includegraphics[width=1\linewidth]{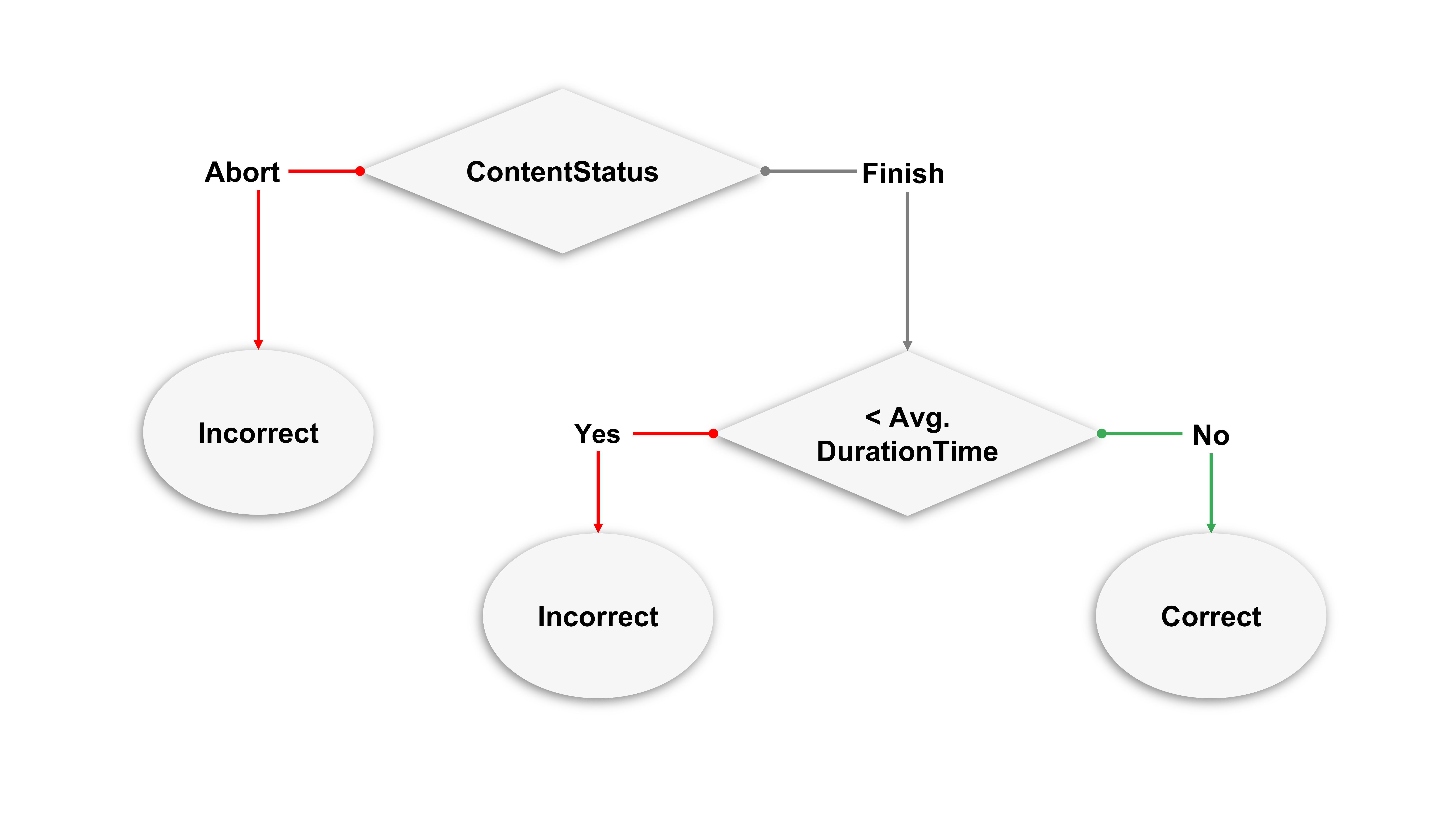}
\caption{
Flowchart for Correctness Determination Based on Game-play Interaction Data
}
\label{fig:flowchart}
\end{figure}

For all game sessions, the session was marked as \textit{Incorrect} if the game was interrupted, which means \textit{Abort} in Figure \ref{fig:flowchart} (e.g., a player exited mid-game). It is important to note that the logs are recorded when a game is completed. Each game consists of multiple stages, and the content status is marked as \textit{Finish} only when all existing stages are completed. If a player only completes part of the stages, the status is recorded as \textit{Abort}. The number of stages in a game can vary from as few as two to more than ten. The number of stages in each game corresponds to the number of dots displayed at the top of the video. For example, as seen in the captured images of three games in Figure \ref{fig:data_info}'s Game Playing Video, the stage counts are 3, 3, and 10, respectively, from left to right.

Based on these log characteristics, the average duration time for each game was calculated from instances where all stages were completed. The correctness determination for games marked as \textit{Finish}, meaning all stages were completed, is explained as follows: if the recorded duration time for a game was longer than the average duration time for the given content, the attempt was considered \textit{Incorrect}. On the other hand, if the duration time was equal to or shorter than the average, the attempt was considered \textit{Correct}. This method allowed us to handle incomplete or skipped attempts across the dataset systematically.

Additionally, extreme outliers were removed from the dataset. Expressly, 397 instances exceeding 30 minutes were excluded to ensure data integrity. After this cleaning process, the final dataset comprises 15,032 users and 7,783,466 problem-solving events.

\subsection{Statistics}
% 여기에는 도희님이 작성

The finalized dataset (see Figure \ref{fig:data_gen_flow} \textit{right} and Table \ref{tb:comparison}) provides a comprehensive overview of player interactions and learning outcomes. Below is a breakdown of key statistics:

\begin{itemize}
    \item \textbf{Total users}: 15,032
    \item \textbf{Total problem-solving events}: 7,783,466
    \item \textbf{Subjects}: Indonesian (IND), Malaysian (MAY), Mathematics (MAT), English (ENG)
    \item \textbf{Categories (KC)}: 28 across all subjects
    \item \textbf{ContentID (Questions)}: 182 across all subjects
\end{itemize}

Figure \ref{fig:plot} provides visualizations that help to understand key aspects of user interactions, such as sequence length, correct answer ratio, learning time, and sequence time, across different subjects. In some visualizations, a log scale facilitates a more straightforward interpretation of the data.

The distribution of students' interaction sequence length shows that 74.2\% of student sequences contain 100 to 1000 interactions, highlighting that most students have substantial interaction sequences. Differences in sequence length by subject reveal that only the Malaysian subject has an average sequence length below 100, indicating that students in this subject tend to solve fewer problems than students in other subjects.

The distribution of correct answer ratios across students' responses shows an average correct answer ratio of 45.19\%. This even distribution suggests that students' responses are balanced between correct and incorrect answers, regardless of the subjects, indicating a balanced learning process.

The focus on students' learning time demonstrates that most students accumulate approximately 10 hours of pure problem-solving time. The distribution of learning time by subject aligns with the findings regarding the Malaysian subject, where students not only complete fewer problems but also spend less cumulative time on problem-solving compared to other subjects.

The distribution of sequence times, which captures the intervals between consecutive problem-solving logs from the app, reveals that a significant portion of logs (over 7 million) occurs within 10 minutes, reflecting frequent and continuous user interaction. The insights provided by sequence times exceeding 10 minutes are significant, as they reveal longer intervals of app usage, with the most common gap being between 1 day and 1 week, suggesting periodic re-engagement patterns with the app.

\section{The ES-KT-24 Benchmark}

\subsection{Experiment Setting}

We comprehensively evaluated various KT models on the ES-KT-24 dataset for our benchmark. We employed a standard 5-fold cross-validation approach to ensure robust performance estimation. All experiments were run multiple times to account for variability, with results being reported as mean values and standard deviations.

\subsection{Baseline Models}

We compared a wide range of KT models, categorized into two main types: traditional DKT models and LKT models \cite{lee2024language}. The DKT models included DKT \cite{piech2015deep}, DKT+ \cite{LS2018_Yeung_DKTP}, DKVMN \cite{zhang2017dynamic}, ATKT \cite{guo2021enhancing}, SAKT \cite{pandey2019self}, and SimpleKT \cite{liusimplekt}. The LKT models included various pre-trained language models: BERT \cite{devlin-etal-2019-bert}, RoBERTa \cite{zhuang-etal-2021-robustly}, ELECTRA \cite{clark2020electra}, ERNIE-2.0 \cite{Sun_Wang_Li_Feng_Tian_Wu_Wang_2020}, ALBERT \cite{lan2019albert}, DistilBERT \cite{sanh2019distilbert}, and DeBERTa-v3 \cite{hedebertav3}.

\subsection{Performance Analysis}

Our experimental results reveal significant insights into the performance of different KT approaches on the ES-KT-24 dataset. Area Under the Curve (AUC) and Accuracy (ACC) were used as the evaluation metrics.

Among the DKT models, SimpleKT demonstrated the best performance with an AUC of 0.7366 and ACC of 0.6709, significantly outperforming traditional approaches. ATKT also showed strong results with an AUC of 0.7100 and ACC of 0.6521. Interestingly, some models like AKT and GKT failed to converge on our dataset, resulting in zero scores.

The LKT models generally outperformed their DKT counterparts, highlighting the effectiveness of leveraging pre-trained language models for KT tasks. RoBERTa achieved the highest performance among all models with an AUC of 0.7348 and ACC of 0.6691, closely followed by ERNIE-2.0 and DeBERTa-v3. Even the worst-performing LKT model (ALBERT) outperformed most DKT models, emphasizing the potential of language model-based approaches in this domain.

The consistent and robust performance across different LKT models is worth noting, with all achieving AUC scores above 0.72 and ACC scores above 0.65. This consistency suggests that the language model-based approach is practical and reliable across various architectures.

The superior performance of LKT models can be attributed to their ability to capture semantic information from the textual content of questions and concepts, which is particularly beneficial in our game-based learning scenario. However, the strong showing of SimpleKT among DKT models indicates that well-designed traditional approaches can still be competitive.

These results underscore the immense potential of language model-based approaches in KT, especially in contexts with rich textual information. They also highlight the value of our ES-KT-24 dataset in advancing research in game-based KT and multimodal learning analytics, paving the way for exciting future developments in the field.

\begin{table}[]
\centering
\small
\begin{tabular}{clll}
\toprule
Type & \multicolumn{1}{c}{Models} & \multicolumn{1}{c}{AUC} & \multicolumn{1}{c}{ACC} \\
\toprule
DKT  & DKT                        & 0.6824±0.0004           & 0.6335±0.0005           \\
DKT  & DKT+                       & 0.6829±0.0007           & 0.6340±0.0007           \\
DKT  & DKVMN                      & 0.6821±0.0005           & 0.6345±0.0002           \\
DKT  & ATKT                       & 0.7100±0.0141           & 0.6521±0.0092           \\
DKT  & SAKT                       & 0.6848±0.0009           & 0.6361±0.0006           \\
DKT  & SimpleKT                   & 0.7366±0.0006           & 0.6709±0.0007           \\
LKT  & BERT                       & 0.7287±0.0010           & 0.6645±0.0012           \\
LKT  & RoBERTa                    & 0.7348±0.0022           & 0.6691±0.0016           \\
LKT  & ELECTRA                    & 0.7303±0.0017           & 0.6660±0.0018           \\
LKT  & ERNIE-2.0                  & 0.7325±0.0014           & 0.6673±0.0012           \\
LKT  & ALBERT                     & 0.7231±0.0021           & 0.6588±0.0014           \\
LKT  & DistilBERT                 & 0.7279±0.0018           & 0.6639±0.0013           \\
LKT  & DeBERTa-v3                 & 0.7326±0.0026           & 0.6670±0.0029 \\
\toprule
\end{tabular}
\caption{Performance of KT models on ES-KT-24. SimpleKT shows the best performance, followed by LKT-RoBERTa.}
\label{tb:performance}
\end{table}

\section{Additional Data Usage for Educational Research}

The ES-KT-24 dataset, with its rich game-based learning data and multimodal information, offers numerous opportunities for educational researchers beyond traditional knowledge tracing. Here are several potential research directions that leverage the unique aspects of our dataset:

\begin{itemize}
    \item \textbf{Game Difficulty Classification}: Researchers can develop models to classify game difficulty levels by analyzing the correlation between gameplay videos and students' correct answer ratios. This could lead to more accurate difficulty scaling in educational games and adaptive learning systems.
    
    \item \textbf{Feature Impact Analysis}: By examining the various features within the games, researchers can identify which elements have the most significant impact on students' performance. This analysis could inform the design of more effective educational games and learning materials.
    
    \item \textbf{Generative Game Design}: Researchers could use gameplay videos to explore the development of AI models capable of generating similar educational games. This approach could lead to the rapid prototyping of new educational games tailored to specific learning objectives.
    
    \item \textbf{Multimodal Learning Analytics}: The combination of gameplay videos, audio, and performance data enables researchers to conduct in-depth multimodal learning analytics. This could reveal insights into how different modes of interaction affect learning outcomes.
    
    \item \textbf{Engagement and Performance Correlation}: Researchers can investigate the relationship between student engagement (as observed in the gameplay videos) and their performance, potentially uncovering new strategies for increasing student motivation and learning efficiency.
    
    \item \textbf{Cross-cultural Learning Patterns}: With data spanning multiple languages (Indonesian, Malaysian, English) and subjects, researchers can explore cross-cultural learning patterns and how they might inform localized educational strategies.
    
    \item \textbf{Temporal Learning Dynamics}: The dataset's temporal information allows for studying how learning patterns evolve, potentially leading to a more nuanced understanding of knowledge retention and optimal spacing for learning sessions.
    
\end{itemize}

\section{Conclusion}

The ES-KT-24 dataset significantly advances KT research, offering a unique combination of game-based learning data and multimodal information. Our comprehensive evaluation of various KT models on this dataset reveals the potential of language model-based approaches in capturing the nuances of student learning in interactive, game-based environments. The performance of LKT models, particularly RoBERTa, underscores the value of leveraging pre-trained language models for KT tasks. However, the strong showing of SimpleKT among traditional DKT models indicates that well-designed conventional approaches remain competitive. These findings highlight the importance of continued research into LKT and DKT methods. Beyond KT, ES-KT-24 opens up numerous avenues for educational research, including game difficulty classification, multimodal learning analytics, and cross-cultural learning pattern analysis.

\section{Limitation}

The ES-KT-24 dataset, while innovative, has several limitations. Firstly, using duration time as the sole indicator of answer correctness may oversimplify student interactions. Future iterations should implement more nuanced classification criteria during data collection. Secondly, the gameplay videos are from researchers rather than actual students, potentially limiting the dataset's ecological validity. Including videos of children playing would provide more authentic representations of learning behaviors.

Additionally, expanding the depth and breadth of log data would enable more comprehensive learning analytics studies. Lastly, while this dataset lays the groundwork for multimodal knowledge tracing, it does not offer solutions for multimodal KT models. This gap highlights a critical area for future research to develop models effectively utilizing multifaceted educational data.

\section{Ethical Consideration}

In this study, we prioritized ethical considerations in several key areas. Firstly, all data was pre-processed to ensure student anonymity and prevent personal identification. We also used gameplay videos recorded by researchers rather than actual students to further protect privacy and avoid ethical concerns.

In the paper preparation process, we utilized Claude and GPT for paraphrasing to enhance readability, strictly limiting their use to improving linguistic quality rather than generating content.

We employed GPT-4o and Whisper for data generation. It is important to note that, by OpenAI's guidelines, the data generated using these tools should be used solely for research purposes and not for enhancing other generative models.

These measures are a testament to our unwavering commitment to maintaining ethical standards in data handling, privacy protection, and the responsible use of AI tools in research.

% \bibliography{main}

% \input{8_appendix}

\end{document}